\documentclass{amsart}
\usepackage{epsf}
\def\Bbb{\mathbb}
\def\BT{\mbox{$\Bbb T$}} 
\def\BC{\mbox{$\Bbb C$}} \def\BP{\mbox{$\Bbb P$}}
\def\BZ{\mbox{$\Bbb Z$}} 

\newtheorem{theorem}{Theorem}[section]

\theoremstyle{definition}
\newtheorem{definition}[theorem]{Definition}
\newtheorem{conjecture}[theorem]{Conjecture}
\newtheorem{example}[theorem]{Example}

\theoremstyle{remark}

\numberwithin{equation}{section}

\begin{document}

\title[D-branes on CY manifolds: a view from the
helix]{D-branes and Vector Bundles on Calabi-Yau Manifolds: a view from
the Helix}

\author[S. Govindarajan]{Suresh Govindarajan}
\address{Department of Physics, Indian Institute of Technology, Madras,
Chennai 600036 INDIA}
\email{suresh@chaos.iitm.ernet.in}
\author{T. Jayaraman}
\address{The Institute of Mathematical Sciences, C.I.T. Campus,
Taramani, Chennai 600113 INDIA}
\email{jayaram@imsc.ernet.in}

\date{May 2001}

\begin{abstract}
We review some recent results on D-branes on Calabi-Yau (CY) manifolds. 
We show the existence of structures (helices and quivers) which enable 
one to make
statements about large families of D-branes in various phases
of the Gauged Linear Sigma Model (GLSM) associated with the CY
manifold. A comparison of the quivers of two phases leads to the prediction
that certain D-brane configurations will decay as one moves
across  phases. We discuss how boundary fermions
can be used to realise various D-brane
configurations associated with coherent sheaves in the GLSM with
boundary. 

This is based on the talk presented by S.G. at Strings 2001,
Mumbai. 
\end{abstract}

\maketitle

\section{Introduction}

This talk is  a summary of recent work done by us on D-branes in
Calabi-Yau manifolds\cite{lsmone,helices,sheaves}. Following the seminal
work of Polchinski\cite{pol}, there has been enormous progress in our
understanding of both BPS and non-BPS branes in flat space. The
situation for the case when spacetime is, say $M^4\times X$,
where X is a Calabi-Yau three-fold is more complicated. Most of our
understanding is limited to special regions in their moduli
space such as the {\em large-volume limit} and {\em Gepner points} where
one uses techniques specially adapted to the situation. 
Unlike the case of toroidal compactifications, these correspond to fewer
unbroken supersymmetries and thus fewer constraints follow. For example,
the BPS condition leaves open the possibility of walls of
marginal stability in the moduli space, where a D-brane can decay.  
It is thus useful to understand the behaviour of D-branes away from 
the special points in regions where stringy corrections are important.

The moduli space of Calabi-Yau manifolds is the direct product of
the K\"ahler moduli space and the complex moduli space. At special
points in these moduli spaces, there are different descriptions of
D-branes. For the quintic, two special points in the K\"ahler moduli
space are the {\em Gepner point} and the {\em large volume point}(see
figure 1). 
\begin{figure}[ht]
\begin{center}
\leavevmode\epsfysize=4cm \epsfbox{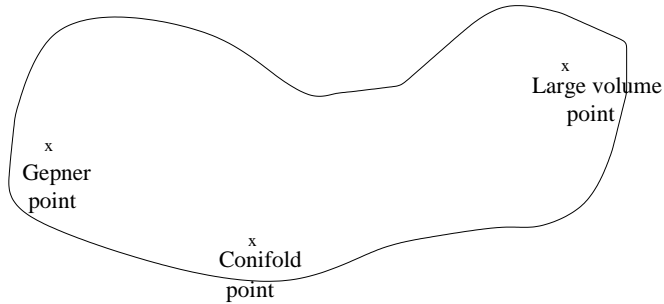}
\end{center}
\caption{K\"ahler moduli space for the quintic}
\end{figure}
In the large volume limit where one has a nice geometric description of
the Calabi-Yau manifold, D-branes which preserve A-type supersymmetry
wrap {\em special Lagrangian submanifolds} while those that preserve B-type
supersymmetry wrap {\em holomorphic cycles} of the Calabi-Yau
space\cite{ooy}.
More generally, including the gauge fields on the brane, we end up with
B-branes being related to coherent sheaves.  We will only consider
B-branes in this talk.

At the Gepner point, one is
in a non-geometric phase and the compactification is described by a
suitable tensor product of ${\mathcal N}=2$ minimal models. The quintic
is given by tensoring five copies of the $k=3$ minimal model.
D-branes are
described, for example, by constructing boundary states satisfying
Cardy's consistency condition\cite{cardy}. A large family of such states have been
constructed using a prescription due to Recknagel and Schomerus\cite{RS}. 
We will
refer to these states as the RS boundary states. 
\begin{example}
For the quintic, the
B-type RS states are described by the following labels\cite{quintic}
$$
|L_1,L_2,L_3,L_4,L_5;M\rangle
$$
where $L_i=0,1$ and $M=0,1,2,3,4$. The $M$ label reflects a quantum
$\BZ_5$ symmetry. The five boundary states given by $\sum_iL_i=0$ are
{\em rigid} i.e., they have no moduli associated with them.
\end{example}

The natural question to ask is what geometric objects in the
large-volume limit correspond to the RS boundary states. The answer
depends on the path taken from the Gepner point to the large volume
point since B-branes can be transformed under monodromy around singular
points\cite{quintic,diacgom}. 
Further, not all B-branes at the Gepner point will be stable
objects in the large-volume limit. This question was studied in
the case of the quintic\cite{quintic}, where the large-volume charges of the RS
boundary states were obtained. Subsequently, in \cite{dfr2}, it was shown
that the five $\sum_i L_i=0$ states are given by the restriction of the bundles
$\Omega^p(p)$ on $\BP^4$ to the quintic hypersurface\footnote{$\Omega$ is the
cotangent bundle to $\BP^4$ and $\Omega^p(p)=\wedge^p \Omega \otimes
{\mathcal O}(p)$.}. It was also conjectured that all $\sum_i L_i\neq 0$ RS
states are bound states of the five $\sum_i L_i=0$. It is rather easy
to verify this at the level of charges. 

The method used in establishing the relationship between RS boundary states
and the large-volume bundles makes use of mirror symmetry and is rather
cumbersome. {\em Is there a more straightforward relationship which does
not make use of mirror symmetry?} In this talk, we will see the
existence of structures which enable us to make this correspondence in a
simpler and more transparent manner.

\section{The Gauged Linear Sigma Model}

The Gauged Linear Sigma Model (GLSM) introduced by Witten provides a
field-theoretic description of the moduli spaces associated with
Calabi-Yau manifolds\cite{wittenphases}. In this description, the Gepner point 
and the
large volume point are seen to be suitable limit points in {\em phases}
of the GLSM. The field content of the GLSM 
are: $\Phi_i$ are $(2,2)$
chiral supermultiplets with charge $Q_i$ with components
$(\phi_i,\psi_{\pm i},F_i)$; $P$ is a chiral supermultiplet with
charge $Q_p = -\sum_i Q_i$ and components $(p,\psi_{\pm p},F_p)$
and an abelian vector multiplet $V$ whose field strength is part of a
twisted chiral superfield $\Sigma$ with components
$(\sigma,\lambda_\pm,D,v_{01})$. The Fayet-Iliopoulos D-term
and theta term are described by a single complex parameter
$t=\frac{\theta}{2\pi} + i r$. We will consider two situations:
\begin{itemize}
\item[(i)] Let us assume that there is no
superpotential for the chiral superfields.
When $r\gg0$, the low energy theory is the the non-linear
sigma model associated with the total space of a line-bundle ${\mathcal
O}(Q_p)$ over  the weighted projective space $X=\BP^{Q_1,\cdots,Q_n}$.
This is a non-compact Calabi-Yau manifold. In the limit when $r\ll0$,
one is in the orbifold phase. The low-energy theory is described by
the orbifold $\BC^n/\BZ_{|Q_p|}$.
\item[(ii)] Suppose we have a superpotential $W=PG(\Phi)$
where $G$ is a polynomial of degree $|Q_p|$. Now in the limit when
$r\gg0$, the low-energy theory is given by the non-linear sigma model
associated with the Calabi-Yau manifold $M$ given by the hypersurface $G=0$
in $X=\BP^{Q_1,\cdots,Q_n}$. The field $p$ has zero vev and its
fluctuations are massive. This is the {\em Calabi-Yau phase}. In the
limit $r\ll0$, one obtains an Landau-Ginzburg (LG)
orbifold as the low-energy theory.  This the {\em LG phase}.
The infrared fixed points of this ultraviolet theory are supposed to be
the large volume and Gepner points respectively.
\end{itemize}
In the following, we shall use $X$ to represent the ambient weighted
projective space in which the Calabi-Yau manifold $M$ is an
hypersurface.

Since the RS boundary states occur in the Gepner point, it is of
interest to see how they are realised in the LG orbifold. It was argued
in \cite{stt} that the only possible boundary conditions on individual LG
fields are Dirichlet and take the form $\phi_i=0$. This implies that
{\em the B-type RS states are branes localised at the orbifold singularity!}.
In the case, when there is no superpotential, such branes are called 
{\em fractional branes}(see \cite{diacgom} and references therein). 
This is indeed the first clue that RS states
must be related in some form to D-branes associated with the orbifold
$\BC^n/\BZ_{|Q_p|}$. We now proceed to discuss D-branes associated with
orbifolds.

\section{D-branes on orbifolds and Quivers}

Building on the work of Douglas and Moore\cite{douglasmoore}, 
it was soon realised that the
various D-branes associated with the orbifold $\BC^n/\Gamma$, where
$\Gamma$ is a discrete group are associated with representations of a 
quiver, the {\em McKay Quiver}\cite{dgm}. The vertices of the quiver are in
one-to-one correspondence with the fractional branes discussed in the
previous section. 
The field content of the worldvolume gauge theory of the
D-brane associated with the quiver is encoded as follows: The positive
integers $n_i$ lead to a gauge group $\prod_i U(n_i)$ and each arrow
corresponds to bi-fundamental scalar fields (as well as their
supersymmetric partners). 
\begin{example}
Consider $\BC^5/\BZ_5$, which is the
orbifold associated with the quintic example. The quiver (see figure 2)
has five vertices associated with the five fractional branes. It
is in agreement with the fact that there are five $\sum_i L_i=0$ 
RS boundary states. 
\begin{figure}[ht]
\begin{center}
\leavevmode\epsfysize=4cm \epsfbox{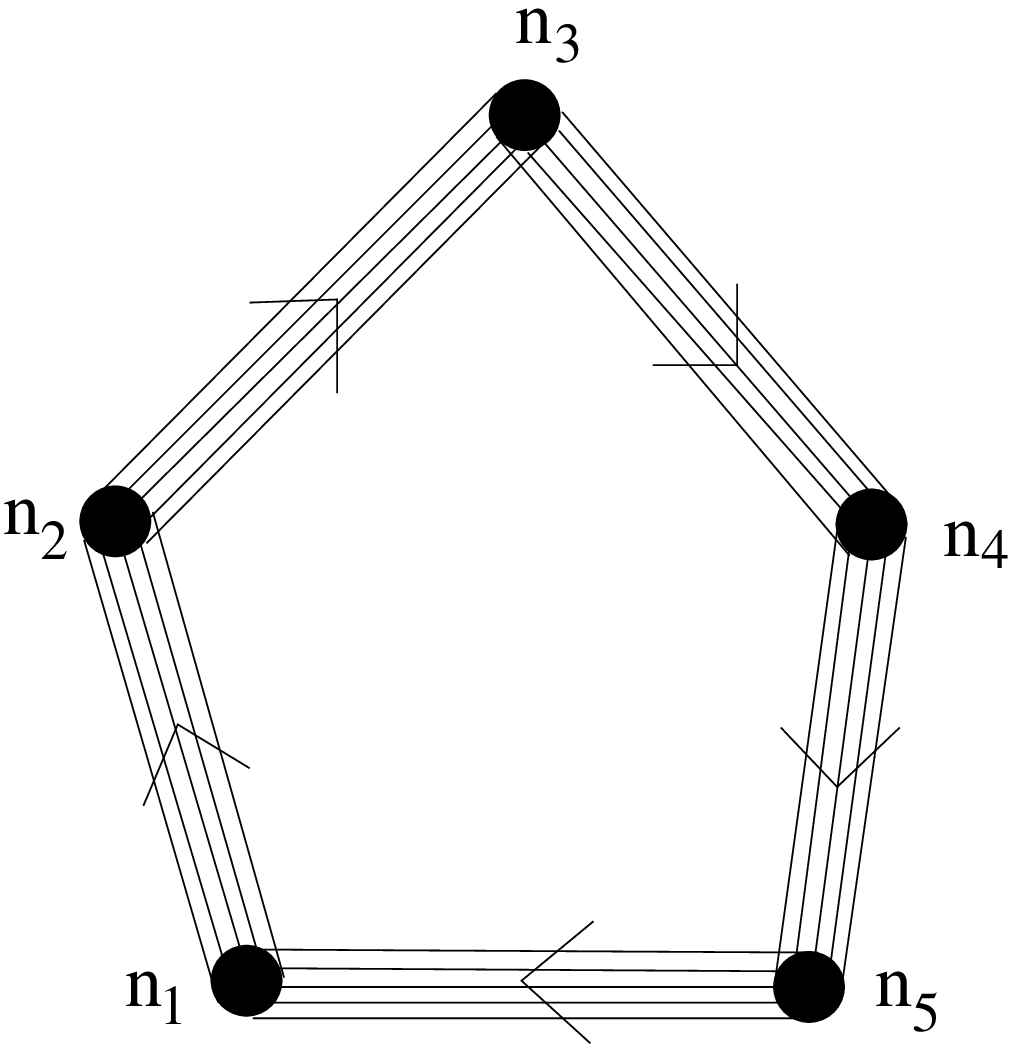}
\end{center}
\caption{The McKay quiver for $\BC^5/\BZ_5$}
\end{figure}
\end{example}

Given an orbifold singularity, one might resolve the singularity by
blowing up. It is not known in general, whether there exists crepant
resolutions (i.e., the resolved space is a Calabi-Yau manifold).
of orbifold singularities except for two cases: $\BC^2/\Gamma$
(where $\Gamma$ is a discrete sub-group of $SU(2)$ and $\BC^3/\Gamma$
(where $\Gamma$ is an abelian sub-group of $SU(3)$). In these cases,
one can see (based on the work of Ito and Nakajima\cite{itonak}) 
that the fractional
branes - $\{S^a\}$ - provide a basis for $K_c(X)$.  $K_c(X)$ represents the
K-theory classes for all D-branes/vector bundles
with compact support i.e., those that live on the exceptional divisors
corresponding to the resolution of the singular space $X$. Further,
these are {\em dual} to {\em tautological bundles} on the orbifold
- $\{R_a\}$ -  
which form a basis for $K(X)$, the K-theory classes associated with the
unresolved space. This is one version of the {\em McKay
correspondence}. The duality is 
$$
\langle R_a,S^b\rangle \equiv \chi(R_a,S^b) = \int_X {\rm ch}(R_a^*) \
{\rm ch}(S^b)\ {\rm Td}(X) = {\delta_a}^b\quad.
$$ 

This is reminiscent of what we saw in the LG and CY phases. In the LG
phase, all branes had support only on singularity. Hence, when the
singularity is blown up -- these branes necessarily have support on the
associated exceptional divisors. The $\{S_a\}$ above have the same
property. Further, the {\em fractional branes} form a basis for the
branes near the orbifold point and hence it is natural to identify them
with the $\{S_a\}$.
Using these observations, Douglas and Diaconescu proposed 
the existence of tautological bundles $\{R_a\}$ in more general
situations\cite{DD}. Further, they gave an inverse toric algorithm
to construct the tautological bundles. This approach has two problems:
it is tedious and one does not know how to handle situations where
the ambient space has singularities which the CY hypersurface 
does not inherit.
In next couple of sections, we will see that the $\{R_a\}$
are the {\em foundation of a helix} 
and can be obtained by studying the
large-volume monodromy of the simplest brane of all -- the six-brane.
This is associated with the line bundle ${\mathcal O}$. The special role
played by the line bundle ${\mathcal O}$ is very much in
line with recent K-theoretic considerations based on Sen's study of
non-BPS branes\cite{sen}.  Just as the D9-brane (and its antibrane)
in IIB string theory
generate all other branes in flat spacetime, it is natural to expect
the six-brane (wrapping all of
$M$) to generate all lower branes and hence charges.

\section{Helices and Mutations}

\begin{definition}
A coherent sheaf $E$ on a  variety $X$ (of dimension $n$)
is called {\em exceptional} if
\begin{eqnarray*}
{\rm Ext}^i(E,E)&=&0\quad,\quad i\geq1 \\
{\rm Ext}^0(E,E)&=&\BC\quad,
\end{eqnarray*}
where Ext${}^i(E,F)$ is the sheaf-theoretic generalisation of the cohomology
groups $H^i(X,E^\ast\otimes F)$ for vector bundles $E$ and $F$. 
\end{definition}
The dimension of Ext${}^i(E,F)$ is the number of Ramond
ground states of charge $i$ of an open-string connecting B-branes
associated with coherent sheaves $E$ and $F$ and $\chi(E,F)$ is a Witten
index\cite{HIV}.
\begin{definition}
An ordered collection of exceptional sheaves ${\mathcal E}=(E_1,\ldots,E_k)$
is called a {\em strongly exceptional collection}\footnote{This is
somewhat different from the definition of Rudakov and is the one
used by Bondal\cite{bondal}. An exceptional collection is one for
which the Ext${}^i(E_a,E_b)=0$ for $a>b$.}
if for all $a<b$, one has
\begin{eqnarray*}
{\rm Ext}^i(E_b,E_a) &=&0 \quad,\ i\geq0  \\
{\rm Ext}^i(E_a,E_b) &=&0 \quad,\ i\neq i_0\quad,
\end{eqnarray*}
for some $i_0$ (which is typically zero).
\end{definition}
This implies that there exist Ramond ground states only in the sector
with charge $i_0$. Further, $\chi(E_a,E_b)$ is an upper - triangular
matrix with ones on the diagonal.

New exceptional collections can be generated from old ones by a process
called {\em mutation}. 
\begin{definition}
A {\em left mutation} of an exceptional pair
$(E_a,E_{a+1})$ in an exceptional collection is defined by\footnote{The
process of mutation has been related to brane creation on the mirror
in \cite{HIV}. See also \cite{zaslow}.}
\begin{equation}
L_{a}(E_a,E_{a+1}) = (L_{E_{a}}(E_{a+1}),E_a)
\end{equation}
where we have introduced a new sheaf 
$L_{E_{a}}(E_{a+1})$ which is defined through exact sequences (see 
\cite{rudakov} for details). For example, when Ext${}^0(E_a,E_{a+1})\neq0$
and the (evaluation) map Ext${}^0(E_a,E_{a+1})\otimes E_a \rightarrow E_{a+1}$
is injective, then $L_{E_{a}}(E_{a+1})$ is defined by
\begin{equation}
0 \rightarrow {\rm Ext}^0(E_a,E_{a+1})\otimes E_a \rightarrow E_{a+1} 
\rightarrow L_{E_{a}}(E_{a+1})
\rightarrow 0
\end{equation}
\end{definition}
The Chern characters of the new sheaves are given by
\begin{equation}
{\rm ch}(L_{E_{a}}(E_{a+1}) = {\rm ch}(E_{a+1}) - \chi(E_a,E_{a+1}) {\rm
ch}(E_a) 
\end{equation}
Further, the collection is assumed to be strongly
exceptional (with $i_0=0$) and hence $\chi(E_a,E_{a+1})={\rm
dim~Ext}^0(E_a,E_{a+1})$.

The mutation of a strongly exceptional collection may not continue to be
strongly exceptional.  If the mutated collection is also strongly
exceptional, the mutation is called {\em admissible}. The mutations that
we consider in this paper (in order to generate $S_i$)
are assumed to be admissible though we do not always verify this explicitly.

\begin{definition}
An exceptional collection $(E_i, \ i\in \BZ)$ is called a {\em helix
of period $p$} if for all $s$ the following condition is satisfied:
\end{definition}
All pairs $(E_{s-1},E_s)$,
$(E_{s-2},L^1(E_s)),\ldots,(E_{s-p+1},L^{p-2}(E_s))$ admit left
mutations and $L^{p-1}((E_s))=E_{s-p}$.

Thus, a sequence of $(p-1)$ left mutations of a helix brings one back
to an element of the helix modulo a shift of $p$. Each collection
$(E_i,E_{i+1},\ldots,E_{i+p})$ is called a {\em foundation} of the
helix $\{E_i\}$. Any helix is determined uniquely by any of its
foundations. One can also define the helix using right mutations.
We shall henceforth use the term helix for the foundation of a helix
since we will have no need to distinguish them.
\begin{example}
The collection of line bundles 
${\mathcal R}=\{ {\mathcal O}, {\mathcal O}(1),\ldots,
{\mathcal O}(n)\}$ is the foundation of a helix of period $n$ on
the complex projective space $\BP^n$.
\end{example}
\subsection{Two conjectures and their consequences}
\begin{conjecture}
The large-volume monodromy ($t\rightarrow
t+1)$ on ${\mathcal O}$ (i.e., the bundle which wraps all of $X$) produces
an exceptional collection which is a helix with foundation
${\mathcal R}\equiv\{R_1={\mathcal O},R_2,\ldots,R_p\}$.
\end{conjecture}
\noindent{\bf Remarks}
\begin{enumerate}
\item The period $p$ of the helix reflects a quantum $\BZ_p$ symmetry
associated with $X$.
\item The procedure seems to work for cases when the space $X$ is {\em
not} fully resolved\cite{helices}.
\item In cases where there are many K\"ahler moduli, the
monodromy action $t_i\rightarrow t_i +1$ ($i=1,\ldots,d$) generates a
$d$-dimensional lattice of which the helix is a one-dimensional
lattice\cite{mayr}.
\item This conjecture has been verified in a variety of
examples\cite{helices,mayr,tomas}.
\end{enumerate}

\begin{conjecture}
All exceptional bundles on $X$ are generated by mutations. 
\end{conjecture}
\noindent{\bf Remarks}
\begin{enumerate}
\item There exists a mutated helix with foundation
$\{S^p,\ldots,S^1={\mathcal O}\}$ with $S^a=L^{a-1}(R_a)$ with the
property:
$$
\chi(R_a,S^b) = \delta_a^b\quad,
$$
i.e., the $S^a$ are dual to the $R_a$.
\item The restriction of the $S^a$ to the Calabi-Yau hypersurface gives
the $\sum_i L_i=0$ RS states in all examples that we have considered.
\item For $X=\BP^n$, it is known that ${\mathcal R}$ form a basis for all
sheaves on $\BP^n$. This follows from Beilinson's theorem\cite{beilpn}. 
This suggests
that there exists a generalisation for weighted projective spaces as
well by using methods followed in \cite{gorrud,drezet} for vector
bundles on $\BP^n$. 
Thus, one anticipates that all coherent sheaves on $X$ will
be given as the cohomology of complexes of associated with the
$R_a$ (or $S_a$).
\end{enumerate}
\section{Quivers from Helices}

Given a helix ${\mathcal R}\equiv \{R_1,\cdots,R_p\}$, one can construct
a quiver as follows\cite{zaslow}: Consider a quiver with $p$ vertices with the 
$i-$th vertex associated with $R_i$. Draw dim$[{\rm Hom}(R_i,R_j)]$ arrows
beginning at vertex $i$ and ending at vertex $j$. The relations of the
quiver are the obvious ones. The following example illustrates the
general situation.
\begin{example}
Consider $X=\BP^4$. Here ${\mathcal R}=\{{\mathcal O}, {\mathcal O}(1),
\ldots,{\mathcal O}(4)\}$. 
\begin{itemize}
\item Hom$({\mathcal O}(i),{\mathcal O}(i+1))=$
multiplication by $\phi_i$ ($i=1,\ldots,5$) where $\phi_i$ are
homogeneous coordinates on $\BP^n$.
\item Relations: Consider Hom$({\mathcal O},{\mathcal O}(2))$. One has
the relation $\phi_1 \phi_2 = \phi_2 \phi_1$.
\item The associated quiver called the {\em Beilinson quiver} (see
figure 3). 
\begin{figure}[ht]
\begin{center}
\leavevmode\epsfysize=4cm \epsfbox{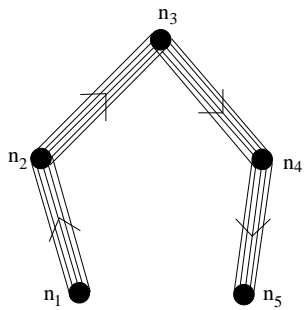}
\end{center}
\caption{The Beilinson quiver for $\BP^4$}
\end{figure}
\item There is a one-to-one correspondence between representations of
this quiver and coherent sheaves (and hence D-branes) on $\BP^4$.
\end{itemize}
\end{example}

It is useful to study the two quivers associated with the two phases of
the GLSM associated with the quintic. If we ignore the superpotential,
we obtain the orbifold $\BC^5/\BZ_5$ in the LG phase and the total space
of the line bundle ${\mathcal O}(-5)$ on $\BP^4$ in the CY phase.
As we have seen the D-branes in the LG phase are associated
with representations of the McKay quiver while D-branes in the CY phase
are associated with representations of the Beilinson quiver. Comparing
the two quivers (see figures 2 and 3), we see that the McKay quiver has
extra arrows connecting vertex $5$ to vertex $1$. If these arrows
were present in the
CY phase, we would require  maps of negative degree. One of the
characteristics of the LG phase is that the $p$ field has non-zero
vacuum expectation value (vev). This suggests that in the LG phase,
these Hom's are given by maps of the form $p\phi_i$, there are five in
all which accounts for the extra five arrows in the McKay quiver. 
\begin{conjecture}
All D-brane configurations that make use of the links that disappear
when one goes from the LG to the CY phase should {\em decay}\cite{dfr2}.
\end{conjecture}
In more general examples one has several phases. 
The above picture suggests
that to each phase (more specifically, a limit point)
has its own quiver whose representations give rise to
D-brane configurations. This will enable one to handle complicated
situations where {\em hybrid phases} appear and when there is no
orbifold/LG phase. The disappearance of links between phases gives
us classes of D-branes that are expected to decay.

\begin{example}
Another example which illustrates the scenario is that of the local
model for a flop given by the total space of line bundles 
${\mathcal O}(-1)\oplus{\mathcal O}(-1)$ over $\BP^1$. 
The GLSM has four fields $(\phi_1,\ldots,\phi_4)$  with charges
$(1,1,-1,-1)$. There are two phases: $r\gg0$ where $\phi_3,\phi_4$
have vanishing vevs and $r\ll0$ where $\phi_1$, $\phi_2$ have
vanishing vevs. The flop-transition is given in figure 4.
\begin{figure}[ht]
\begin{center}
\leavevmode\epsfysize=2.5cm \epsfbox{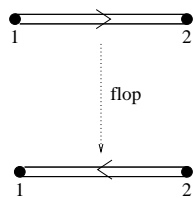}
\end{center}
\caption{The quivers for the flop transition}
\end{figure}
This example also illustrates the fact that flop-transition corresponds
to a {\em change of t-structure} in the derived category naturally
associated with representations of the quiver\cite{partha}.
\end{example}

\section{Coherent Sheaves in the GLSM with Boundary}

It is of interest to see how D-brane configurations associated with
coherent sheaves can be realised in the GLSM with boundary. We will for
simplicity consider the case when a coherent sheaf $E$ is given as the
cohomology of a monad. The more general case is briefly discussed
at the end of this section.
The basic idea is to consider the following complex
(monad) of holomorphic vector bundles  $A$, $B$ and $C$
\begin{equation}
0\rightarrow A \stackrel{a}{\rightarrow} B \stackrel{b}{\rightarrow}
C \rightarrow 0\quad,
\end{equation}
which is exact at $A$ and $C$. 
The holomorphic
vector bundle $$E={\rm ker}\ b/{\rm Im}\ a$$ is the cohomology of the
monad.  

In a field-theoretic realisation, this is implemented as follows. Consider
fermions $\pi_a$ ($a=1,\ldots,{\rm rk}~B)$. The map $a$ is realised as the 
{\em gauge invariance}
$$
\pi_a \sim \pi_a + E_a^i(\phi) \kappa_i\quad,
$$
where $\kappa_i$ are sections of $A$ ($i=1,\ldots,{\rm rk}~A)$. This
gauge-invariance is fixed by the condition(s)
$\overline{E}_a^i \pi_a =0$. The map $b$ is implemented by the {\em
holomorphic constraint}
$$
J_m^a(\phi)\pi_a =0 \quad (m=1,\ldots,{\rm rk}~C)\quad.
$$

In the GLSM, we will be interested in the case where the boundary
preserves the linear combinations of the bulk $(2,2)$ supersymmetry
associated with B-type boundary conditions. In this regard, it is useful
to see how the bulk fields decompose. We obtain: (i) A $(2,2)$ chiral
multiplet $\Phi$ decomposes into a scalar and Fermi chiral multiplet
$(\Phi',\Xi)$ respectively. (ii) A twisted chiral multiplet $\Sigma$
becomes an unconstrained complex multiplet. (iii) The combination
$\widetilde{v}_0 =v_0+ \eta \frac{\sigma + \overline{\sigma}}{\sqrt2}$ 
behaves as the boundary gauge field.

We introduce boundary Fermi multiplets satisfying
\begin{equation}
\overline{\mathcal D} \Pi_a = \sqrt2 \Sigma' E_a(\Phi')
\end{equation}
where $\Sigma'$ is a boundary chiral multiplet. Similarly, the
holomorphic constraints require the addition of a boundary chiral
multiplet. 

So far the story seems similar to $(0,2)$ constructions for vector
bundles in the GLSM. One new ingredient is that {\em large volume
monodromy} of the vector bundles must be implemented correctly.
For example, for the quintic, under $t\rightarrow t+1$, $E\rightarrow
E\otimes {\mathcal O}(1)$. In the monad construction, it is easy to see
that this corresponds to a simple shift in the $U(1)$ charges of the
boundary fermions.

In the GLSM, this is done by the addition of boundary contact terms --
part of which can be fixed in the Non-Linear Sigma Model  (NLSM) limit of the
GLSM. Further, the boundary conditions in the GLSM must have a proper
NLSM limit and requires careful treatment of the fields in the vector
multiplet\cite{stt,unpub}. See \cite{sheaves} for more details.

One consequence of Beilinson's theorem, for say, $\BP^n$, is that all
coherent sheaves arise from complexes of length less than or
equal to $n$ ($n=2$ is a monad).
Thus, in order to describe D-branes associated with complexes of
arbitrary length, one needs to extend the above discussion to this
situation. This requires ones to deal with nested gauge invariances
and more fields. Further, in the GLSM with boundary it is more natural
to consider direct sums of line-bundles rather than vector bundles.
This can also be implemented in the GLSM with boundary provided one
uses  {\em first-order actions} for the bosonic boundary superfields.
This also seems to make the implementation of the large volume monodromy
much simpler\cite{sheaves}.

We illustrate it with an example in $\BP^4$.
Consider ten fermi superfields $\Pi_{[ij]}$ subject to the
constraint ($i,j,k=1,2,\ldots,5$ and $E_{ij}^k = 
\frac12(\phi_i \delta_j^k - \phi_j \delta_i^k)$)
$$
\overline{\mathcal D} \Pi_{[ij]} = \sqrt2  \Sigma_k
E_{[ij]}^k(\Phi')\quad.
$$
where $\Sigma_k$ are five bosonic superfields subject to the constraint
($E_k=\phi_k$)
$$
\overline{\mathcal D} \Sigma_k = \sqrt2 N E_k(\Phi') \quad.
$$
Here we have introduced a chiral fermi superfield $N$. The consistency
condition between the two constraints is
$ \sum_k E_{[ij]}^k\  E_k =0.$
This is the statement that the composition of two consecutive maps in
the following complex vanish.
\begin{equation}
0\rightarrow {\mathcal O}(-2) \stackrel{E_k}{\rightarrow} 
{\mathcal O}(-1)^{\oplus5} \stackrel{E_{[ij]}^k}{\longrightarrow}
{\mathcal O}^{\oplus 10} \rightarrow \BT^2(-2)\rightarrow0
\end{equation}
One can verify that the number of massless fermions are as expected and
are sections of $\BT^2(-2)$. ($\BT$ is the tangent bundle to $\BP^4$.)

\section{Conclusion}

In this talk we have seen a correspondence between
D-branes at special limit points in various phases of CY manifolds
and representations of quivers. Using this, we have seen how
one can predict that certain D-branes will decay as one moves from one
phase to another. However, the precise points at which they decay needs
more detail and is discussed in the talk by Douglas at this conference.
One may wonder if all D-branes on the Calabi-Yau manifold arise from the
quivers that we considered. This clearly cannot be the case since not
all vector bundles on a Calabi-Yau manifold arise as restrictions of
bundles from the ambient variety. This is also related to the fact that
the RS states are only a subset of the all boundary states satisfying
Cardy's consistency condition.
There are other arguments based on
completeness in an axiomatic formulation of string field
theory\cite{lazar} which also lead to similar conclusions.

\noindent{\bf Acknowledgments} We thank T.~Sarkar for useful discussions
and collaboration.

\bibliographystyle{amsalpha}

\end{document}